\documentclass[epj]{webofc}
\usepackage[varg]{txfonts}

\woctitle{International Conference on New Frontiers in Physics 2013}

\newcommand{\detadphi}    {\ensuremath{\Delta\eta\Delta\varphi}}
\newcommand{\pp}    {\ensuremath{pp}}

\begin{document}
\title{Two-particle angular correlations in \pp~collisions recorded with the ALICE detector at the LHC}

\author{Ma{\l}gorzata Janik\inst{1}\fnsep\thanks{\email{majanik@if.pw.edu.pl}} (for the ALICE Collaboration)
}

\institute{Faculty of Physics, Warsaw University of Technology, ul. Koszykowa 75, 00-662 Warszawa, Poland}

\abstract{We report on the studies of two-particle angular correlations measured in proton-proton collisions at a center-of-mass energy of $\sqrt{s}=7$ TeV recorded by ALICE at the LHC. 
Two-particle correlations in relative azimuth ($\Delta\varphi$) and pseudorapidity ($\Delta\eta$) are expected to   exhibit several structures which arise from different physics mechanisms and allow us to study the wide landscape of correlations. 
The results include the dependence of the correlation function on the event multiplicity, the charge combination and species (pions, kaons or protons) of particles in the pair.}

\maketitle

\section{Introduction}
\label{sec:intro}
Measurements of two-particle correlations are a
powerful tool to study the underlying mechanism of particle production 
in collisions of hadrons and nuclei at high energy. Such studies often involve 
the measurement of relative angles $\Delta\varphi$ and $\Delta\eta$, 
where $\Delta\eta=\eta_1 - \eta_2$ is the difference in
pseudorapidity, $\Delta\varphi=\varphi_1 - \varphi_2$ is the difference in
azimuthal angle.  In general, correlation functions are sensitive to different sources i.e. minijets, elliptic flow, Bose-Einstein correlations, 
resonance decays, photon conversions, etc \cite{Janik:2012ya,GraczykowskiIS2013}.  
%The obtained result is the effect of the contributions of all those correlations.
%Each of the correlation sources has its own and unique structure in \detadphi~space. 
Contributions from those correlations create distinctive structures in \detadphi~space. 
The goal of this study is to decompose them and quantify each contribution separately. 
%Of particular interest is the study of such correlations in order to characterize minijets,
%but also any measurable contributions from Bose-Einstein correlations, elliptic flow,
% resonances, photon conversions and others .

Furthermore, in the studies of untriggered \detadphi~correlations in $pp$ collisions we expect only some of the effects that are present in collisions of heavy-ions (e.g. we do not expect elliptic flow), and deepening our knowledge of interactions of small systems can serve as a baseline for the understanding of Pb--Pb collisions.
% in pp we expect only some of the effects and it serves a s a baseline for the understanding of PbP
Moreover, there is a direct relation between $\Delta\eta$ and $\Delta\varphi$ observables and the relative momentum of the pair (the femtoscopic correlations are measured as a function of pair relative momentum) \cite{Janik:2012ya}.

In this paper we present the measurements of angular correlations of non-identified and identified particles (protons, kaons and pions) in \pp~collisions at a center-of-mass collision energy $\sqrt{s}=7$~TeV registered by the ALICE experiment. 

\section{Data analysis}
\subsection{Data selection}
\subsubsection*{Event and track selection}  
A data sample of 530 millions of minimum bias $pp$ collisions at $\sqrt{s}=7\ \rm{TeV}$ from the 2010 run of the LHC was studied. A detailed description of the ALICE detector can be found in Ref. \cite{Aamodt:2008zz}. The main subsystems used for the presented analysis are the Inner Tracking System (ITS), the Time Projection Chamber (TPC) and the Time Of Flight detector (TOF). 

The analysis was performed on primary particles within the acceptance range of $|\eta|<1.0$  and transverse momentum range $p_{\mathrm{T}}>0.12$ GeV/$c$ (for the analysis of non-identified particles) and $p_{\mathrm{T}}>0.1$ GeV/$c$, $p_{\mathrm{T}}>0.3$ GeV/$c$, and  $p_{\mathrm{T}}>0.5$ GeV/$c$  for pions, kaons, and protons respectively.  Tracks were required to have a Distance of Closest Approach (DCA) to the reconstructed collision vertex smaller than 2 cm in the beam direction. In the transverse plane, a  $p_{\mathrm{T}}$  dependent cut was applied and only particles within ($\mathrm{0.018+0.035}p_{\mathrm{T}}^{-1.01}$) cm from the primary vertex were accepted. Furthermore, we applied a cut to remove electron-positron pairs which were produced from photon conversions. It was achieved by removing pairs constructed of one positive and one negative particle that have small polar angle difference ($\Delta\theta<0.008$) and invariant mass $m_{\mathrm{inv}}<0.002$ (close to $0$, the rest mass of the photon). The applied event and track selections are described in more details in Ref. \cite{Janik:2012ya}.
\subsubsection*{Particle Identification}
The particle identification was performed using the combined TPC and TOF detector information. The identification of particles as pions, kaons, and protons was based on the difference (expressed in units of the resolution  $\sigma$) between the measured signal and the expected signal in the TPC and TOF. The  $\sigma_{\mathrm{TPC}}$  is the standard deviation from the TPC Bethe-Bloch parametrization of $\mathrm{d}E/\mathrm{d}x$ energy loss signal and $\sigma_{\mathrm{TOF}}$  is the time-of-flight deviation from the expected time of the analyzed particle for each particle species. Depending on the momentum of the particle $\mathrm{N}_{\sigma,\mathrm{TPC}}$ varies from 2 to 5, and $\mathrm{N}_{\sigma,\mathrm{TOF}}$ from 2 to 3. Particles were efficiently identified (purities above 90\%) up to $p_{\mathrm{T}}$ of 2.1~GeV/$c$ for pions, up to 1.43~GeV/$c$ for kaons and up to 2.6~GeV/$c$ for protons.
\subsubsection*{Monte Carlo models}
The same analysis was performed on Monte Carlo (MC) events generated by two different models -- Pythia Perugia-0 \cite{Sjostrand:2006za,Skands:2010ak} and Phojet \cite{Engel:1995sb}. Moreover, the MC generated events were processed through the reconstruction chain of the ALICE framework in order to simulate the detector response. The particles selected for the analysis had to pass the same set of track and pair cuts as the ALICE collision data. The results do not differ significantly from the ones obtained from the analysis of the generator-only information.

\subsection{Correlation function}
The two-particle experimental correlation function is defined as a function
of $\Delta\eta$ and $\Delta\varphi$  as follows: 
\begin{equation}
\label{eq:CorrelationFuntion}
C(\Delta\eta,\Delta\varphi)=\frac{N_{pairs}^{mixed}}{N_{pairs}^{signal}} \frac{S(\Delta\eta,\Delta\varphi)}{B(\Delta\eta,\Delta\varphi)},
\end{equation}
where $N_{pairs}^{signal}$ is the number of pairs
constructing the signal $S$, and $N_{pairs}^{mixed}$ is the number of
pairs in the background $B$. The signal is determined by counting particle pairs within the same
\detadphi~range in the same event. The background is estimated by applying the event mixing technique. Details on the procedure of constructing the correlation functions are described in Ref. \cite{Janik:2012ya}.

\section{Results}
\label{sec:results}
In this paper we focus on the fitting of $\sqrt{s}=7\ \rm{TeV}$ results and the correlation functions for identified particles. The \detadphi~correlation functions for non-identified particles can be found in Ref. \cite{Janik:2012ya}. In particular the distributions for $pp$ collisions at $\sqrt{s}=0.9\ \rm{TeV}$, $\sqrt{s}=2.76\ \rm{TeV}$ and $\sqrt{s}=7\ \rm{TeV}$ collision energies, different charge combinations (four different sets of particle pairs were analyzed: all possible, positive like-sign, negative like-sign, unlike-sign), dependent on the multiplicity and on pair transverse momentum are presented there. To quantify different effects seen in the those correlation functions a multi-component fit to the distributions was performed.

\subsection{Fitting procedure for non-identified particles}
\label{etaphiresults:fitting}
To analyze the obtained experimental correlation functions
quantitatively we propose a mathematical formula to describe all the observed
correlation features. We have taken each of them into account when constructing the final fitting formula (\ref{eq:F1}).

Minijets, Bose-Einstein (femtoscopic) correlations in like-sign pairs and 
resonances in unlike-sign pairs are
added to the fitting function as a 2-dimensional modified Gaussian
component at the $\Delta\eta = \Delta\varphi = 0$. 

An away-side correlation (wide ridge for  $\Delta\varphi=\pi$) is formed due to the presence of back-to-back jets and momentum conservation.
Particles that were produced in two opposite back-to-back jets have $\Delta\varphi$  close to $\pi$. However, there is no strong dependence on the pseudorapidity of those particles; so, for many events and many jets, $\Delta\eta$ is almost uniform, which results in a ridge for  $\Delta\varphi=\pi$. Simultaneously conservation laws would create similar correlation to back-to-back jets. 
The resulting structure is modeled by a 1-dimensional Gaussian at $\Delta\varphi = \pi$. 

Finally, the soft component (string fragments) represents a longitudinal fragmentation in
unlike-sign pairs and produces a 1-dimensional Gaussian
correlation centered at $\Delta\eta=0$. This 1-dimensional
structure was observed also by other experiments \cite{Alver:2007wy, CMS:2010tua, ATLAS:2012ap} and is consistent with expectations from the low-mass resonance gas model \cite{Eggert:1974ek}, Monte Carlo model of isotropic decay of clusters \cite{Alver:2008aa}, and local charge conservation (charge ordering) in longitudinally fragmenting strings \cite{Porter:2005rc}, described in the string fragmentation model such as the Lund model \cite{Henyey:1973hs,Berger:1974vn,Meunier:1974nj,Michael:1975uf}.

To take all mentioned structures into account, we used the following fitting function: 
%\begin{eqnarray}
\begin{equation}
\label{eq:F1}
F_1=N \left [1+M_{\mathrm{M}} \exp \left(- \left(\frac{\Delta\varphi^2} 
{2\sigma_{\mathrm{M}\varphi}^2}+\frac{\Delta\eta^2} {2\sigma_{\mathrm{M}\eta}^2}
\right)^{e_\mathrm{M}}\right)
+ M_{\mathrm{A}} \exp \left(- \frac{(\Delta\varphi-\pi)^2}
{2\sigma_{\mathrm{A}\varphi}^2}\right)
+ M_{\mathrm{L}} \exp \left(- \frac{\Delta\eta^2} {2\sigma_{\mathrm{L}\eta}^2}
  \right)
  \right ] \left(1 + P \Delta\eta^2\right).
\end{equation}
%\end{eqnarray}
$M_{\mathrm{M}}$, $\sigma_{\mathrm{M}\varphi}$, $\sigma_{\mathrm{M}\eta}$ and $e_{\mathrm{M}}$ are parameters of the  modified Gaussian describing the near-side peak. $M_{\mathrm{A}}$, $\sigma_{\mathrm{A}\varphi}$  are meant to describe
the away-side correlation. $M_{\mathrm{L}}$, $\sigma_{\mathrm{L}\eta}$  accounts for the longitudinal ridge. In addition, $N$ is the overall normalization and $P$ is the
parameter fitted to the background structures observed in the data visible for higher $|\Delta\eta|$ values (accounting for some unwanted effects of the procedure).

\begin{figure}[!ht]
\centering
\includegraphics[width=\textwidth]{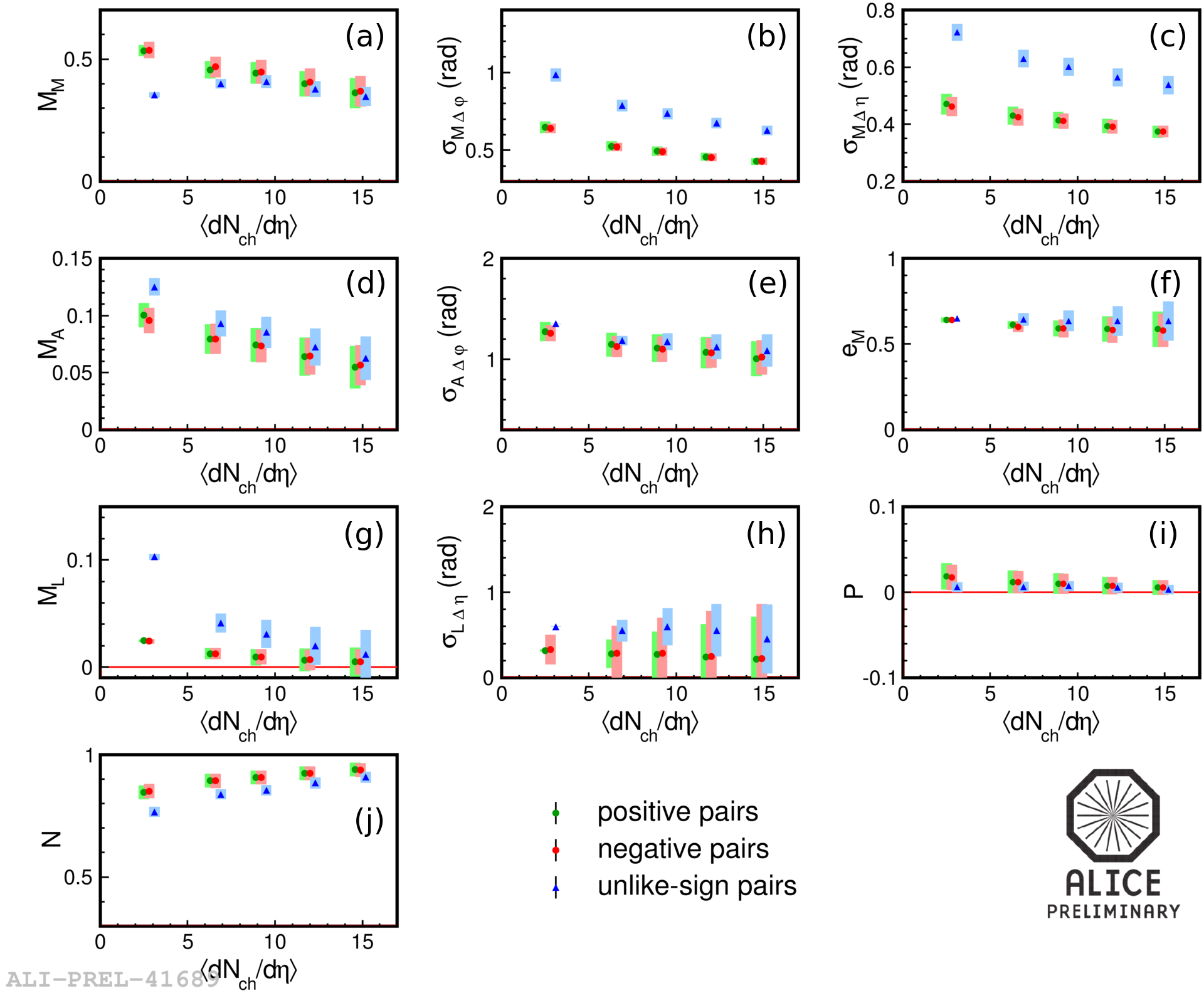}
\caption{Fit parameters for \pp~collisions at
    $\sqrt{s}=7$~TeV as a function of event multiplicity for three
    charge combinations: ``positive'' for two     positive particles,
    ``negative'' for two negative particles and 
    ``unlike-sign'' for pairs of positive and negative particles. No
    selection on pair momentum has been applied. The top row panels show
    the parameters  of the near-side ``minijet'' peak: 
    (a) magnitude $\mathrm{M_M}$, (b) relative azimuthal width $\mathrm{\sigma_{M\sigma\varphi}}$, (c) relative  pseudorapidity width $\mathrm{\sigma_{M\sigma\varphi}}$.
    The next row gives the parameters of the away-side correlation: (d) magnitude $\mathrm{M_A}$,
    (e) relative azimuthal width $\mathrm{\sigma_{A\sigma\varphi}}$, 
    (f) exponent of the near-side peak $\mathrm{e_M}$. 
    The third row shows the longitudinal ridge parameters: 
    (g) magnitude $\mathrm{M_L}$, (h) relative pseudorapidity width $\mathrm{\sigma_{L\sigma\eta}}$, (i) background parabola magnitude P.
    The last row panel (j) shows the overall normalization N.}
\label{fig:detadphiFitHBT} 
\end{figure}

%The correlation functions obtained from $\sqrt{s}=7\ \rm{TeV}$ \pp~collisions 
%and presented in Ref. \cite{Janik:2012ya,JanikPresentationICNFP2013} were
%fitted using the formula (\ref{eq:F1}). 
Figure \ref{fig:detadphiFitHBT}
presents the dependence of the parameters of the full fit, for three
charge combinations, as a function of the charged particle multiplicity density $\mathrm{d}N_{\mathrm{ch}}/\mathrm{d}\eta$ (number of charged particles per pseudorapidity unit), with no
selection on the pair momentum. Below we briefly describe the
trends seen in the plot.

The panels in the top row show the parameters of the near-side
component. We observe a distinct difference between the same and
opposite charge pairs; the latter show a significantly wider peak,
which reflects a stronger correlation effect.  
We also see that this peak has an exponent smaller than
unity, meaning that the shape is more peaked than a Gaussian. We
observe that both the magnitude and the width of the peak decrease
with multiplicity. Such behavior is expected  due to the construction of the correlation function for correlations coming from minijets. The higher the particle multiplicity, the smaller the correlation per pair.

The second row of plots shows the behavior of the away-side ridge. Its
magnitude also decreases with multiplicity as well as its width. We
have tried fits with non-unity values of the exponent, similarly to the near-side peak case; however, the results were always very close to unity, so we kept the unmodified 1-dimensional Gaussian for this fit. 

The third row shows the parameters of the longitudinal ridge, which
is only significant in the lowest multiplicity bins and
for opposite-charge pairs. For the higher multiplicity bins it disappears.

In summary, 
%both structures (near-side peak and away-side ridge)
%reflect mainly the correlations coming from the minijets: 
the magnitude and width of all the correlation structures
decreases with multiplicity as expected. The correlation is stronger for the near-side and away-side structures
for opposite charge pairs, which is consistent with the assumption that they are mainly influenced by minijets. 
The longitudinal ridge is significant only for unlike-sign pairs, supporting the hypotheses of fragmenting strings and low-mass resonances.

\subsection{Correlation functions for identified particles}

\begin{figure}[h]
  \centering
  \includegraphics[width=\textwidth]{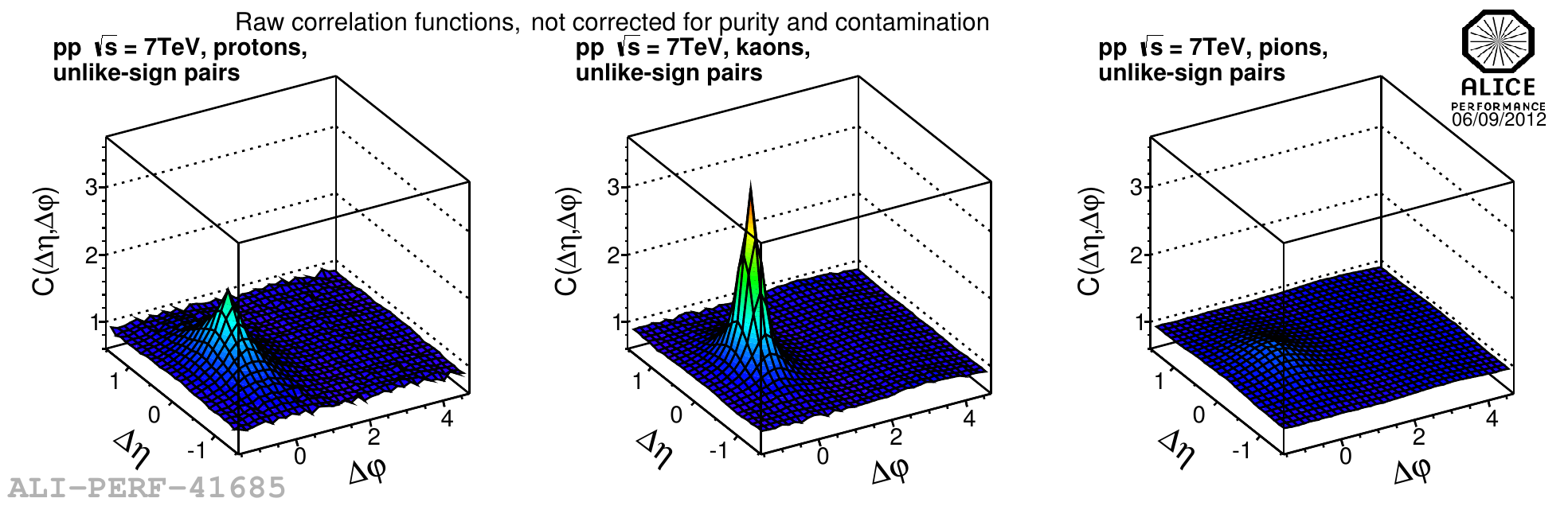}
  \caption{Correlation functions for unlike-sign pairs of protons, kaons and pions. The distributions are not corrected for efficiency, purity and contamination.}
  \label{Fig:corrMergedMixed}
\end{figure}

\begin{figure}[h]
  \centering
  \includegraphics[width=\textwidth]{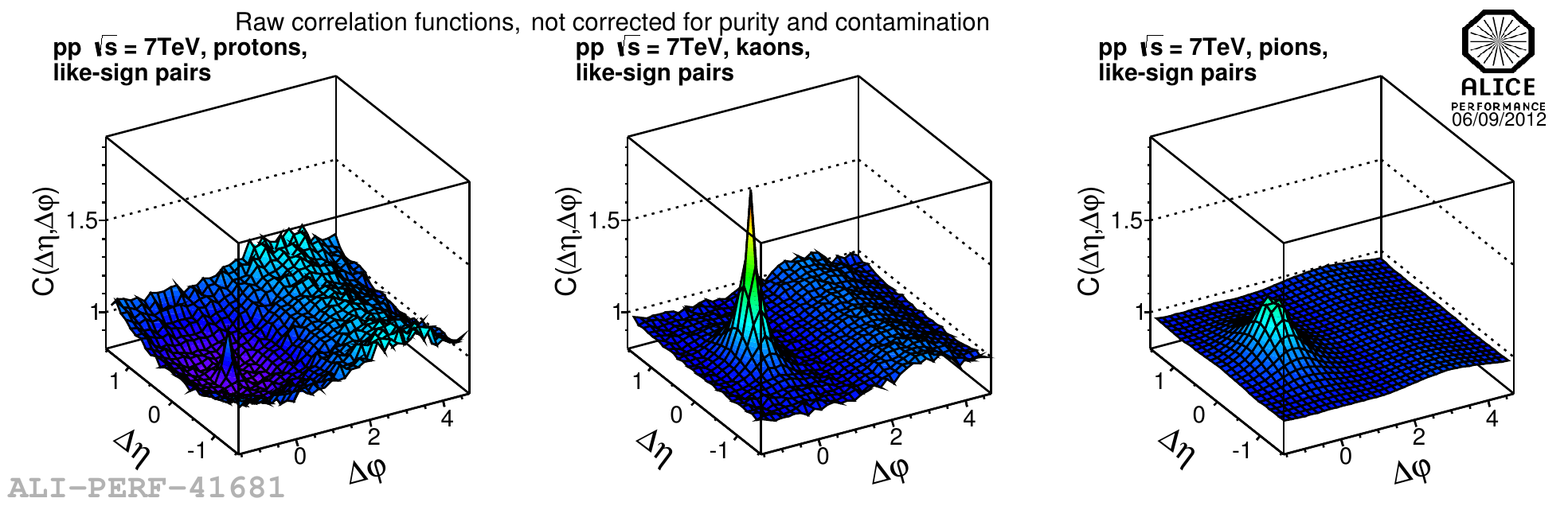}
  \caption{Correlation functions for like-sign pairs of protons, kaons and pions. The distributions are not corrected for efficiency, purity and contamination.}
  \label{Fig:corrMergedSame}
\end{figure}

Figures~\ref{Fig:corrMergedMixed} and \ref{Fig:corrMergedSame} present the correlation functions for different particle types (protons, kaons, pions) for unlike-sign pairs and like-sign pairs. We can observe significantly different shapes of the correlation functions for different particle species. The correlation function is the strongest (the near-side peak is most distinct) for kaon pairs and significantly lower for pions and protons, both for like- and unlike-sign pairs. Moreover, a surprising behavior can be seen for like-sign proton pairs, where a wide anti-correlation dip with minimum at  $(\Delta\eta,\Delta\varphi) = (0,0)$ can be observed.

\begin{table}[ht]
\centering
\caption{The conservation laws that are present in correlations for different types of particles.}
\label{tab:conservLaws}
\begin{tabular}{ccccc}
\hline
particles	&momentum	&charge	&strangeness	&baryon number\\\hline
pions	& \checkmark	& \checkmark	&	        &\\
kaons	& \checkmark	& \checkmark	& \checkmark	&\\
protons	& \checkmark	& \checkmark	&              	& \checkmark\\\hline
\end{tabular}
%\vspace*{5cm}  % with the correct table height
\end{table}

To explain these results we propose the following hypothesis. For the production of different types of particles (pions, kaons, protons) different conservation laws come into play, as summarized in Tab. \ref{tab:conservLaws}. For pions, only momentum and electric charge must be conserved. Charged kaons contain strange quarks; so, they carry strangeness. Therefore, in case of kaons, also the strangeness must be conserved. Protons are baryons; so, the baryon number must be conserved. Since every single collision is an independent system, it must conserve all of these quantities. If those laws are conserved for the whole event we are defining the global conservation laws (i.e. the total strangeness of all produced particles should be zero). Moreover, all the quantum numbers must be conserved for each parton fragmentation separately (i.e. the total strangeness of all particles in the single jet must be zero)~\cite{Aihara:1986fy,Althoff:1984ut,Buskulic:1994ny,Abreu:1997mp,Acton:1993ux,Muller:1999yg}. We would define local conservation laws, as conservation laws that are obeyed for each parton fragmentation. We test at which scale this conservation takes place, since they should influence the shape of the correlation function, especially in small systems. The study of different particle species in $pp$ events shall allow to determine that.  
All of those must be taken into account while interpreting the results of the angular correlations analysis.

For the unlike-sign pairs one can see that the near-side peak centered at $(\Delta\eta,\Delta\varphi) = (0,0)$ has a different magnitude for different particle species. It is most prominent (the correlation is strongest) for kaons, lower for protons, and the weakest for pions. This observation is consistent with our assumption that the conservation laws play a main role in defining the shape of the correlation function.
The basic solution that would ensure conservation of all the laws is the production of particle-antiparticle pair. Moreover, it is always energetically most favorable to compensate particles with their antiparticles. The strength of the correlation depends on the energetic price of the alternative solution. If the alternative solution is ``cheap'' then the unlike-sign pairs (pairs constructed of particle and its antiparticle) will be less correlated, than for the ``expensive'' alternative solutions, where the correlation will be stronger:
(1) for pions the alternative solution is just another opposite-charge particle, and the cost of such a solution is relatively low; so, the correlation is weakest,
(2) for protons another anti-baryon (charged, or neutral plus additional charged particle) would have to be produced. Such solutions are less probable (the energetic price is higher) than in $\pi^{+}\pi^{-}$ case which results in stronger near-side correlation,
(3) for kaons, which carry the strange quark, the strangeness must be conserved; so, the alternative solution would be at least a $\Lambda$ particle together with another baryon. The energetic price of such solution is very high, therefore the creation of $K^{+}K^{-}$ pair is much more favorable and the correlation is the strongest.

For like-sign pairs different mechanisms play a role determining the shape of the correlation function. Producing two identical particles is no longer the ``cheapest solution''. For correlations of such particles their masses play a significant role. The pions are light particles (with rest mass close to 0.139 GeV/$c$), kaons are heavier (with a rest mass about 0.949 GeV/$c$) and protons are the heaviest (with rest mass close to 0.938 GeV/$c$) of the analyzed particles.
In the case of like-sign pairs also the femtoscopic effects which increase the correlation have to be taken into account.

In the case of kaons and pions we can see the prominent near-side peak in the correlation function. In the case of protons a large dip near the  $(\Delta\eta,\Delta\varphi) = (0,0)$ is present. It is due to the fact, that producing two protons, two identical heavy particles, in roughly the same direction, is less probable, than the similar production of some lighter particles. 
In the case of two identical protons produced in a similar direction automatically we would have to produce also two anti-baryons (i.e. two antiprotons) in the opposite direction; so, another two heavy particles. The cost of such a solution is high; so, such cases are rare and therefore we see a dip in the correlation function.

\begin{figure}[h]
  \centering
  \includegraphics[width=\textwidth]{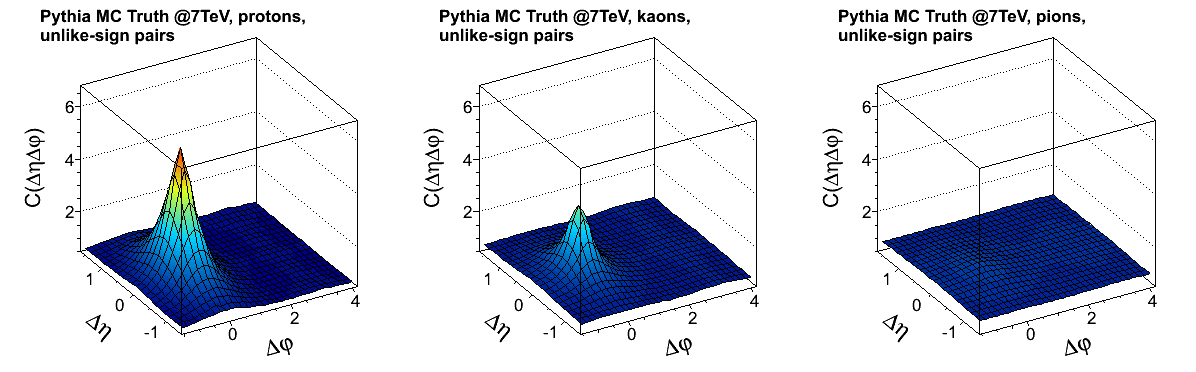}
  \caption{Correlation functions for Pythia Perugia-0 for unlike-sign pairs of protons, kaons and pions. The detector effects were not taken into account.}
  \label{Fig:PythiaMixed}
\end{figure}

\begin{figure}[h]
  \centering
  \includegraphics[width=\textwidth]{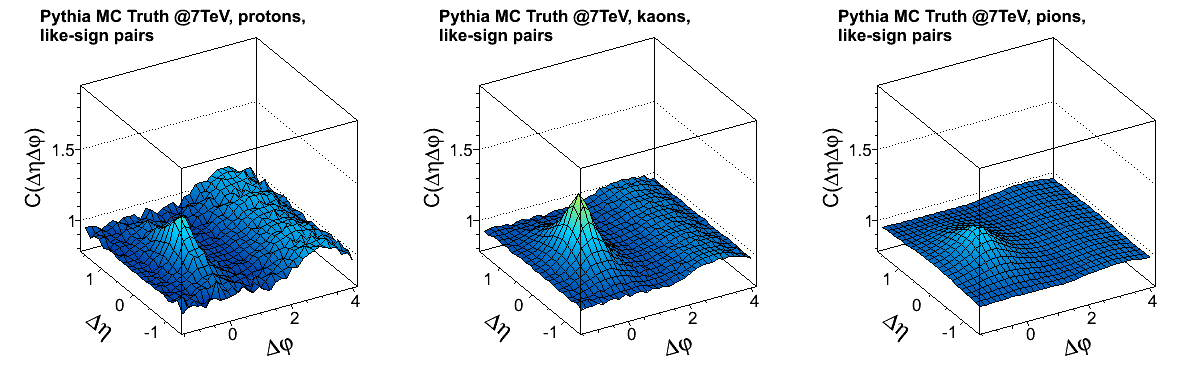}
  \caption{Correlation functions for Pythia Perugia-0 for like-sign pairs of protons, kaons and pions. }
  \label{Fig:PythiaSame}
\end{figure}

The correlation functions obtained from Pythia Perugia-0 are presented in Fig. \ref{Fig:PythiaMixed} for unlike-sign pairs and in
Fig. \ref{Fig:PythiaSame} for like-sign pairs of protons, kaons and pions. 
Correlation functions obtained from Phojet are very similar to Pythia. 
There are significant differences between the results obtained from MC models and from the ALICE collision data (figures \ref{Fig:corrMergedMixed} and \ref{Fig:corrMergedSame}).
The results show that in the studied MC models the strongest correlation exists between unlike-sign pairs of protons, not for kaons as is the case of the results of the analysis of collision data. 
Furthermore, there is a significant near-side peak instead of anti-correlation dip for the like-sign proton pairs. The results suggests that conservation laws, especially the baryon conservation law, are not modeled properly in the tested MC generators. 

%In studied MC models (Pythia, Phojet)  the baryon number is conserved only globally, and the differences between obtained correlation functions from model and from collision data are striking.
%On the other hand the strangeness is conserved for each fragmentation separately, therefore qualitatively results from the model reflect the features seen in the data (i.e. the near-side peak is very prominent for kaons).

Summarizing, conservation laws seem to play a significant role determining the shape of \detadphi~correlation functions. The character  of the like-sign proton correlation suggests that the baryon conservation law plays a major role at a scale significantly smaller than the whole event. This seems not to be precisely modeled in the studied Monte Carlo generators (Pythia, Phojet).

\section{Conclusions}
\label{sec:conclusions}
The \detadphi~angular correlations for non-identified and identified particles (pions, kaons, protons) have been measured in \pp~collisions at $\sqrt{s}=7$~TeV with the ALICE experiment. 
 The analysis for different combinations of charges of particles in the pair (like-sign and unlike-sign) and for different multiplicity ranges was performed. The results were quantified using a fitting procedure. 
The dependence on the multiplicity of the \detadphi~correlation function shows a decrease of the correlation with increasing multiplicity.
The different shapes of the correlation function for each particle type are consistent with the assumption that the conservation laws influence differently the results for pions, kaons and protons.

\section*{Acknowledgments}
\label{sec:acknowledgments}
This work has been financed by the Polish National Science Centre under decisions no. DEC-2011/01/B/ST2/03483, DEC-2012/05/N/ST2/02757, and by the European Union in the framework of European Social Fund through the Warsaw University of Technology Development Programme, realized by Center for Advanced Studies.

\bibliography{bibliography}

\begin{thebibliography}{22}

\bibitem{Janik:2012ya}
M.~Janik, PoS \textbf{WPCF2011}, 026 (2011)

\bibitem{GraczykowskiIS2013}
{\L}.~Graczykowski (ALICE Collaboration), to be published in Nucl. Phys. A
  (2014)

\bibitem{Aamodt:2008zz}
K.~Aamodt et~al. (ALICE Collaboration), JINST \textbf{3}, S08002 (2008)

\bibitem{Sjostrand:2006za}
T.~Sjostrand, S.~Mrenna, P.Z. Skands, JHEP \textbf{0605}, 026 (2006)

\bibitem{Skands:2010ak}
P.Z. Skands, Phys.Rev. \textbf{D82}, 074018 (2010)

\bibitem{Engel:1995sb}
R.~Engel, J.~Ranft, S.~Roesler, Phys.Rev. \textbf{D52}, 1459 (1995)

\bibitem{Alver:2007wy}
B.~Alver et~al. (PHOBOS Collaboration), Phys.Rev. \textbf{C75}, 054913 (2007)

\bibitem{CMS:2010tua}
{CMS Collaboration}, CMS-PAS-QCD-10-002  (2010)

\bibitem{ATLAS:2012ap}
G.~Aad et~al. (ATLAS Collaboration), JHEP \textbf{1205}, 157 (2012)

\bibitem{Eggert:1974ek}
K.~Eggert, H.~Frenzel, W.~Thome, B.~Betev, P.~Darriulat et~al., Nucl.Phys.
  \textbf{B86}, 201 (1975)

\bibitem{Alver:2008aa}
B.~Alver et~al. (PHOBOS Collaboration), Phys.Rev. \textbf{C81}, 024904 (2010)

\bibitem{Porter:2005rc}
R.~Porter, T.~Trainor (STAR Collaboration), Acta Phys.Polon. \textbf{B36}, 353
  (2005)

\bibitem{Henyey:1973hs}
F.~Henyey, Phys.Lett. \textbf{B45}, 469 (1973)

\bibitem{Berger:1974vn}
E.L. Berger, Nucl.Phys. \textbf{B85}, 61 (1975)

\bibitem{Meunier:1974nj}
J.~Meunier, G.~Plaut, Nucl.Phys. \textbf{B87}, 74 (1975)

\bibitem{Michael:1975uf}
C.~Michael, Nucl.Phys. \textbf{B103}, 296 (1976)

\bibitem{Aihara:1986fy}
H.~Aihara et~al. (TPC/Two Gamma Collaboration), Phys.Rev.Lett. \textbf{57},
  3140 (1986)

\bibitem{Althoff:1984ut}
M.~Althoff et~al. (TASSO Collaboration), Phys.Lett. \textbf{B139}, 126 (1984)

\bibitem{Buskulic:1994ny}
D.~Buskulic et~al. (ALEPH Collaboration), Z.Phys. \textbf{C64}, 361 (1994)

\bibitem{Abreu:1997mp}
P.~Abreu et~al. (DELPHI Collaboration), Phys.Lett. \textbf{B416}, 247 (1998)

\bibitem{Acton:1993ux}
P.~Acton et~al. (OPAL Collaboration), Phys.Lett. \textbf{B305}, 415 (1993)

\bibitem{Muller:1999yg}
D.~Muller et~al. (SLD Collaboration), Nucl.Phys.Proc.Suppl. \textbf{86}, 7
  (2000)

\end{thebibliography}

\end{document}